\documentclass[showpacs,amsmath,amssymb,aps,showkeys,floatfix,prd,a4paper]{revtex4}

\usepackage[dvips]{graphicx}
\usepackage{dcolumn}
\usepackage{bm}
\usepackage{epsfig}
\usepackage{amsfonts}
\usepackage{amssymb,amscd}

\def\lsim{\raise0.3ex\hbox{$<$\kern-0.75em\raise-1.1ex\hbox{$\sim$}}}

\def\gsim{\raise0.3ex\hbox{$>$\kern-0.75em\raise-1.1ex\hbox{$\sim$}}}

\newcommand{\be}{\begin{equation}}

\newcommand{\ee}{\end{equation}}

\def\beq{\begin{equation}}

\def\eeq{\end{equation}}

\def\beqa{\begin{eqnarray}}

\def\eeqa{\end{eqnarray}}

\newcommand{\ba}{\begin{eqnarray}}

\newcommand{\ea}{\end{eqnarray}}

\def\gappeq{\mathrel{\rlap {\raise.5ex\hbox{$>$}}

{\lower.5ex\hbox{$\sim$}}}}

\def\lappeq{\mathrel{\rlap{\raise.5ex\hbox{$<$}}

{\lower.5ex\hbox{$\sim$}}}}

\def\Toprel#1\over#2{\mathrel{\mathop{#2}\limits^{#1}}}

\begin{document}

\title{Dark photons from pions produced  in  ultraperipheral $PbPb$ collisions}
\author{V.P. Gon\c{c}alves$^1$ and  B.D.  Moreira$^{2}$}
\affiliation{$^1$ High and Medium Energy Group, Instituto de F\'{\i}sica e Matem\'atica,  Universidade Federal de Pelotas (UFPel)\\
Caixa Postal 354,  96010-900, Pelotas, RS, Brazil.\\
$^2$ Departamento de F\'isica, Universidade do Estado de Santa Catarina, 89219-710 Joinville, SC, Brazil.  \\
}

\begin{abstract}
In this letter we propose the search of  dark photons in the decay of pions produced  by $\gamma \gamma$ interactions in ultraperipheral $PbPb$ collisions. The cross section is estimated considering  an accurate treatment for the absorptive corrections and for the nuclear form factor. Predictions for the event rates are presented considering the expected luminosities for the LHC,  High -- Luminosity LHC and  High -- Energy LHC as well as for the Future Circular Collider. Our results indicate that a future experimental analysis of the pion production in ultraperipheral heavy ion collisions can be useful to probe the dark photon production and constrain its properties.
\end{abstract}



\maketitle

\vspace{1cm}

One of the more challenge problems in Particle Physics today is the description of the composition and nature of Dark Matter (DM), which constitutes at least the 85 \% of the mass of our Universe (For a recent review see e.g. Ref. \cite{Bertone:2016nfn}). 
During the last decades, several possible extensions of the Standard Model (SM)  were proposed aiming to explain the DM abundance \cite{Beacham:2019nyx}. In particular, 
proposed U(1) extensions of the Standard Model gauge group SU(3) x SU(2) x U(1) have raised attention in recent years. In some of these  U(1) extensions a new, light messenger particle $A^{\prime}$, the dark photon, is predicted, with the dark photon field coupling to the hidden sector as well as to the electromagnetic current of the Standard Model  by kinetic mixing. Such coupling allows for a search for the dark photon in laboratory experiments exploring the
electromagnetic interaction  \cite{Boveia:2018yeb,Filippi:2020kii}.
Various experimental programs have been started to search for the $A^{\prime}$ boson (See e.g. Refs. 
\cite{Batley:2015lha,Anastasi:2016ktq,Lees:2017lec,Adrian:2018scb,Aaij:2017rft,Aaij:2019bvg,CortinaGil:2019nuo}). In particular, several experiments have focused in the study of meson decays as an alternative for probing the dark photon production and constrain its properties 
\cite{Batley:2015lha,CortinaGil:2019nuo}.
Such studies are strongly motivated by the fact that meson decays have already been studied for a long time to understand the 
structure of mesons and, consequently, large data sets of meson decays exist from
various experiments.  One possibility is the study of the pion decay, searching for a deviation from the SM pion decay. In the
SM, the neutral pion $\pi^0$ decays dominantly into two photons. However, the decay into one real and a virtual photon from which an electron-positron pair is produced still gives a large signal, which allows to search for a peak associated to the dark photon production in the  smooth
$e^+ e^-$ invariant-mass spectrum predicted by the Standard Model. Dark photon search in meson decays provides some of the most stringent constraints on the properties of dark photons. Recently, the NA62 Collaboration at the CERN SPS has reported results 
\cite{CortinaGil:2019nuo} of a search for $\pi^0$ decays to a photon and a dark photon, improving previous limits in the dark photon mass $m_{A^{\prime}}$ and coupling strength $\epsilon^2$. However, large regions of the ($m_{A^{\prime}}$, $\epsilon^2$) plane are still unexplorated.

\begin{figure}
\centerline{\psfig{figure=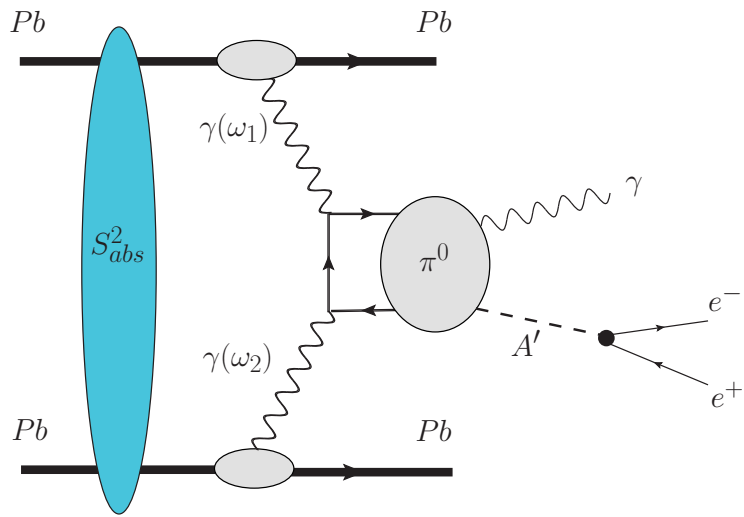,width=10cm}}
\caption{Dark photons from pions produced  by $\gamma \gamma$ interactions in ultraperipheral $PbPb$ collisions.}
\label{fig:diagram}
\end{figure}

High -- energy hadronic collisions at the LHC and future hadronic colliders are an important alternative to 
search for the dark photon and Dark Matter candidates \cite{Boveia:2018yeb,Bruce:2018yzs}. Recent theoretical studies have proposed distinct strategies to probe the dark photon properties and constrain its mass and coupling 
\cite{Ilten:2015hya,Curtin:2014cca,Ilten:2016tkc,Ilten:2018crw,Bauer:2018onh}. Some them already been used by the LHCb Collaboration, which has searched by the dark photon in  $A^{\prime} \rightarrow \mu^+ \mu^-$ decays \cite{Aaij:2019bvg}. Our goal in this paper is investigate the dark photon production in the decay of neutral pions produced in ultraperipheral heavy ion collisions (UPHIC), which are characterized by an impact parameter $b$ greater than the sum of the radius of the colliding  nuclei \cite{epa,upc1,upc2,upc3,upc4,upc5,upc6,upc7,upc8,upc9}. In these collisions, strong electromagnetic fields are generated, which  enhance by a factor $Z^4$ ($\approx 45 \times 10^6$ for $PbPb$ collisions)  the   $\pi^0$  production by two -- photon interactions  \cite{bert,bert_nos}, where $Z$ is number of protons in the nucleus. Moreover, the final state generated by the $\pi^0$ decay is produced in a very clean environment, where pileup is absent and that is characterized by  two intact nuclei and  two rapidity gaps, i.e. empty regions  in pseudo-rapidity that separate the intact very forward nuclei from the final state. Such aspects strongly reduce the probability of mis-identifying the primary vertex and, consequently, reduce possible backgrounds for displaced vertex signatures related to the dark production and decay. In this exploratory study, we will present, for the first time, the total cross section and event rates for the dark photon production  in ultraperipheral $PbPb$ collisions (For related studies see e.g. Refs. 
\cite{Goncalves:2010dw,Goncalves:2015oua,Knapen:2016moh,Baldenegro:2019whq,Beresford:2019gww,Beresford:2018pbt,Drewes:2018xma,Drewes:2019vjy,Coelho:2020saz,Dyndal:2020yen}). In particular, we will consider the process represented in Fig. \ref{fig:diagram}, where the  dark photon decays into a $e^+ e^-$  pair. Our calculations will be performed  using the  equivalent photon approximation, which has been successfully applied for the calculation of the dilepton production in ultraperipheral heavy ion collisions, considering a realistic description of the nuclear form factor and for the treatment of the absorptive corrections \cite{nos_dilepton}. We will present  predictions for the next run of the LHC \cite{hl_lhc}, as well as for the energies of the  High -- Energy LHC ($\sqrt{s} = 10.6$ TeV) \cite{he_lhc} and Future Circular Collider ($\sqrt{s} = 39$ TeV) \cite{fcc}.

The cross section for the process represented in Fig. \ref{fig:diagram} can be factorized in terms of the cross section for the $\pi^0$ production in UPHIC and the branchings for the  transitions $\pi^0 \rightarrow A^{\prime} \gamma$
and  $A^{\prime} \rightarrow e^+ e^-$, as follows:
\begin{eqnarray}
\sigma \left(Pb Pb \rightarrow Pb \otimes \gamma e^+ e^- \otimes Pb; s \right) = \sigma \left(Pb Pb \rightarrow Pb \otimes \pi^0 \otimes Pb \right) \times 
\mbox{BR}(\pi^0 \rightarrow A^{\prime} \gamma) \times \mbox{BR}(A^{\prime} \rightarrow e^+ e^-) \,\,,
\end{eqnarray}
where $\sqrt{s}$ is center - of - mass energy of the $PbPb$ collision and $\otimes$ characterizes a rapidity gap in the final state.
The production of neutral pions ($\pi^0$) in  ultraperipheral $PbPb$ collisions can be described using the equivalent photon approximation \cite{epa}, with the cross section being given by
\begin{eqnarray}
\sigma \left(Pb Pb \rightarrow Pb \otimes \pi^0 \otimes Pb;s \right)   
&=& \int \mbox{d}^{2} {\mathbf b_{1}}
\mbox{d}^{2} {\mathbf b_{2}} 
\mbox{d}W 
\mbox{d}Y \frac{W}{2} \, \hat{\sigma}\left(\gamma \gamma \rightarrow \pi^0 ; 
W \right )  N\left(\omega_{1},{\mathbf b_{1}}  \right )
 N\left(\omega_{2},{\mathbf b_{2}}  \right ) S^2_{abs}({\mathbf b})  
  \,\,\, .
\label{cross-sec-2}
\end{eqnarray}
where $Y$ is the rapidity of the pion in the final state and $W = \sqrt{4 \omega_1 \omega_2}$ is the invariant mass of the $\gamma \gamma$ system. The photon energies $\omega_i$ can be expressed in terms of $W$ and $Y$ as follows: 
\begin{eqnarray}
\omega_1 = \frac{W}{2} e^Y \,\,\,\,\mbox{and}\,\,\,\,\omega_2 = \frac{W}{2} e^{-Y} \,\,\,.
\label{ome}
\end{eqnarray}
Moreover, $N(\omega_i, {\mathbf b}_i)$ is the equivalent photon spectrum  
of photons with energy $\omega_i$ at a transverse distance ${\mathbf b}_i$  from the center of nucleus, defined in the plane transverse to the trajectory, which  can be expressed in terms of the  charge form factor $F(q)$. 
In our analysis, we will consider the realistic form factor, which corresponds to the Wood - Saxon distribution and is the Fourier transform of the charge density of the nucleus, constrained by the experimental data (For a detailed discussion see Ref.  \cite{nos_dilepton}). The factor $S^2_{abs}({\mathbf b})$ depends on the impact parameter ${\mathbf b}$ of the $PbPb$ collision and  is denoted the absorptive  factor, which excludes the overlap between the colliding nuclei and allows to take into account only ultraperipheral collisions. As in Refs. \cite{nos_dilepton,nos_muonium}, we will estimate this quantity using the Glauber approach proposed in Ref. \cite{Baltz_Klein}. Finally, the cross section for the $\gamma \gamma \rightarrow \pi^0$ process can be calculated using the Low formula \cite{Low}, being given by
\begin{eqnarray}
 \hat{\sigma}_{\gamma \gamma \rightarrow \pi^0}(\omega_{1},\omega_{2}) = 
8\pi^{2} \frac{\Gamma_{\pi^0 \rightarrow \gamma \gamma}}{m_{\pi^0}} 
\delta(4\omega_{1}\omega_{2} - m_{\pi^0}^{2}) \, ,
\label{Low_cs}
\end{eqnarray}
where $\Gamma_{\pi^0 \rightarrow \gamma \gamma}$ is the two-photon decay width and $m_{\pi^0}$ is the pion mass.

The branching ratio for the transition $\pi^0 \rightarrow A^{\prime} \gamma$ can be estimated considering that the $A^{\prime}$ field  interacts with the SM photon through a kinetic -- mixing Lagrangian, being given by (see e.g. Ref. \cite{Ilten:2015hya})
\begin{eqnarray}
\mbox{BR}(\pi^0 \rightarrow A^{\prime} \gamma) = 2 \epsilon^2 \left(1 - \frac{m_{A^{\prime}}^2}{m_{\pi^0}^2}\right)^3 \times \mbox{BR}(\pi^0 \rightarrow \gamma \gamma)\,\,.
\end{eqnarray}
 Finally, assuming that the dark photon decay only in SM particles, we will take  $\mbox{BR}(A^{\prime} \rightarrow e^+ e^-) = 1$, which is a reasonable approximation in the mass range $2 m_e \ll  m_{A^{\prime}} < m_{\pi^0}$ considered in our analysis.

 \begin{figure}[t]
\begin{tabular}{cc}
\hspace{-1cm}
 (a) Central detector ($|Y| \le 2.0$)   &   (b) Forward detector ($ 2.0 \le Y \le 4.5$) \\
\hspace{-1.2cm} {\psfig{figure=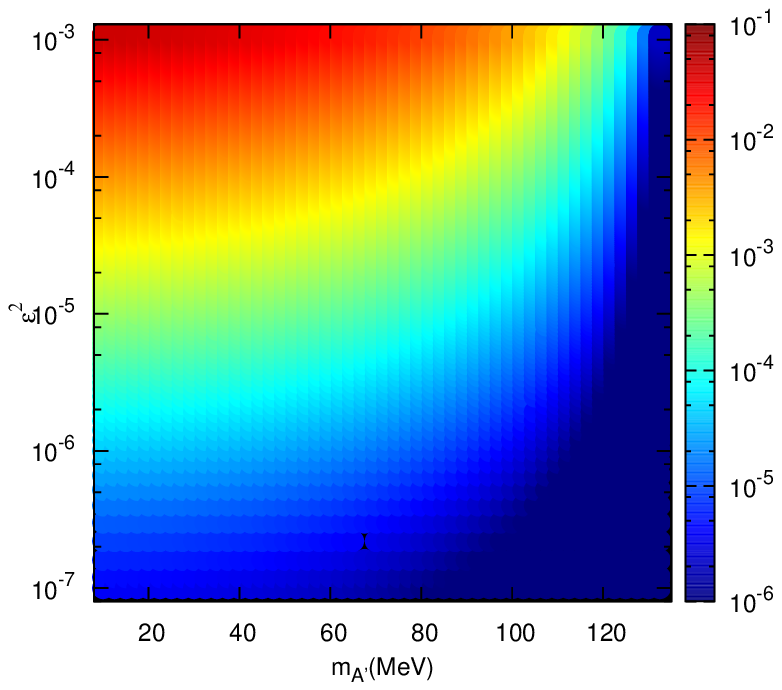,width=10.2cm}} & 
{\psfig{figure=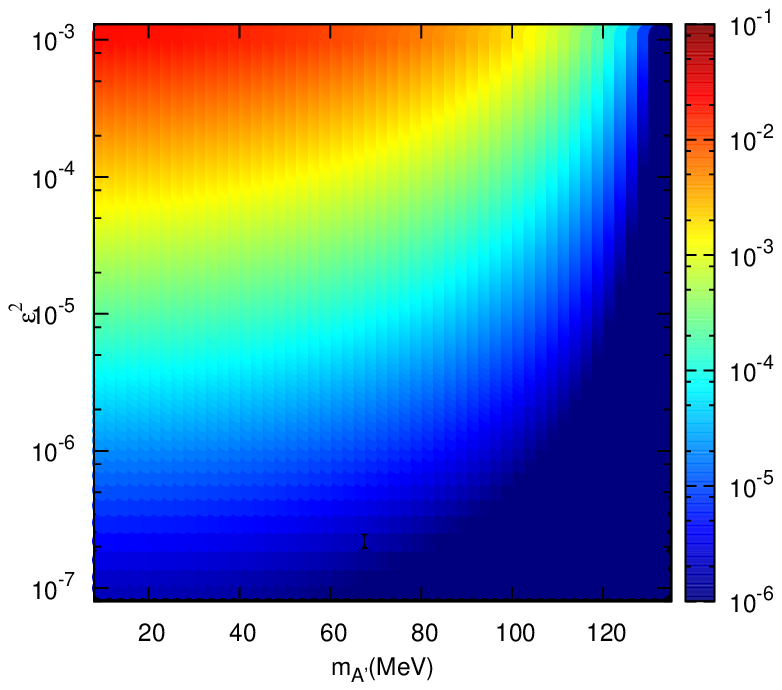,width=10.2cm}} \\ 
\end{tabular}                                                                                                                       
\caption{Cross sections (in mb) for the dark photon production in $PbPb$ collisions at $\sqrt{s} = 5.5$ TeV considering different values for the coupling $\epsilon^2$ and dark photon mass  $m_{A^{\prime}}$ as well as the rapidity range  probed by  (a) central  and (b) forward detectors.}
\label{fig:crosssections}
\end{figure}

Initially, let's estimate the cross sections for $PbPb$ collisions at $\sqrt{s} = 5.5$ TeV and the typical rapidity ranges probed by  central detectors  ($|Y| \le 2.0$), as e.g. by the ALICE, ATLAS and CMS detectors, as well as by a forward detector ($ 2.0 \le Y \le 4.5$), as the LHCb one. As our predictions are dependent on the values for the coupling $\epsilon^2$ and for the dark photon mass  $m_{A^{\prime}}$, we present in Fig. \ref{fig:crosssections} the 
two -- dimensional ($m_{A^{\prime}},\epsilon^2$) distribution for the cross sections, which are given in mb. One has that the cross sections are larger with the increasing of $\epsilon^2$ and the decreasing of $m_{A^{\prime}}$. In particular, for $\epsilon^2 = 10^{-3}$ and $m_{A^{\prime}} = 10$ MeV (upper left corner) we predict $\sigma = 3.9 \times 10^{-2} \,\, (2.1 \times 10^{-2})$ mb for a central (forward) detector. On the other hand, for 
$\epsilon^2 = 10^{-7}$, the cross sections are  reduced by three orders of magnitude. It is important to compare these values with those  predicted for the main background, associated to the process where we have the direct transition $\pi^0 \rightarrow \gamma e^+ e^-$, without the presence of a dark photon in the intermediate state. For this case we have estimated that $\sigma_{bcg} = 2.4 \times 10^{-1} \,\, (1.2 \times 10^{-1})$ mb for a central (forward) detector. Therefore, we predict a  signal/background (S/B) ratio in the range $10^{-1}$ -- $10^{-5}$, depending on the values for the coupling and dark  photon mass. However, it is important to emphasize that such ratio is expected to be strongly enhanced if the displaced vertex and/or resonance strategies are implemented \cite{Ilten:2015hya}.  Moreover, assuming that the cross section for the direct $\gamma e^+ e^-$ prooduction is well known and  constrained by data, such background could be removed, allowing to access those events  associated to the dark photon. Surely, such aspect deserves a more detailed analysis, which we intend to perform in a near future.

\begin{figure}[t]
\begin{tabular}{cccc}
\hspace{-2.2cm} LHC   &  HL -- LHC & HE -- LHC & FCC \\
\hspace{-2.2cm} {\psfig{figure=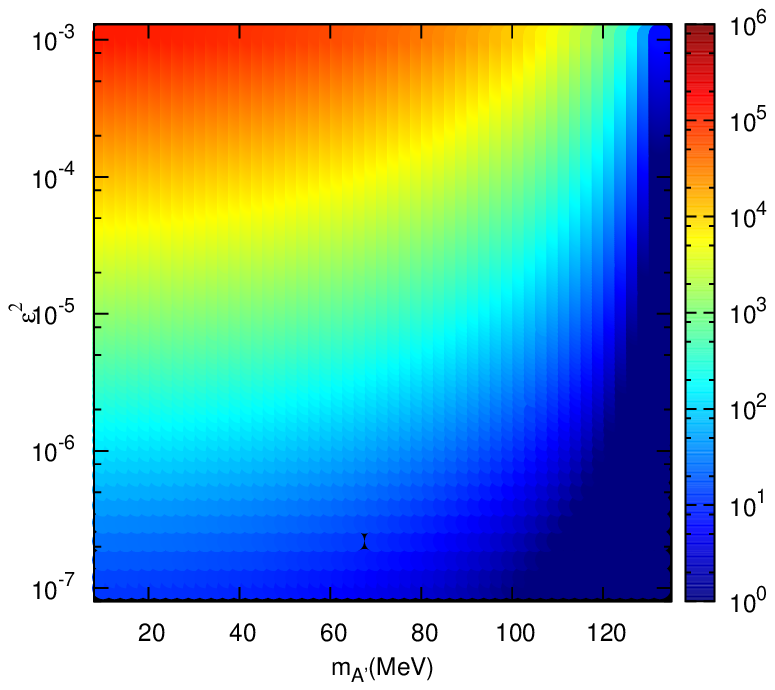,width=5.3cm}} &  {\psfig{figure=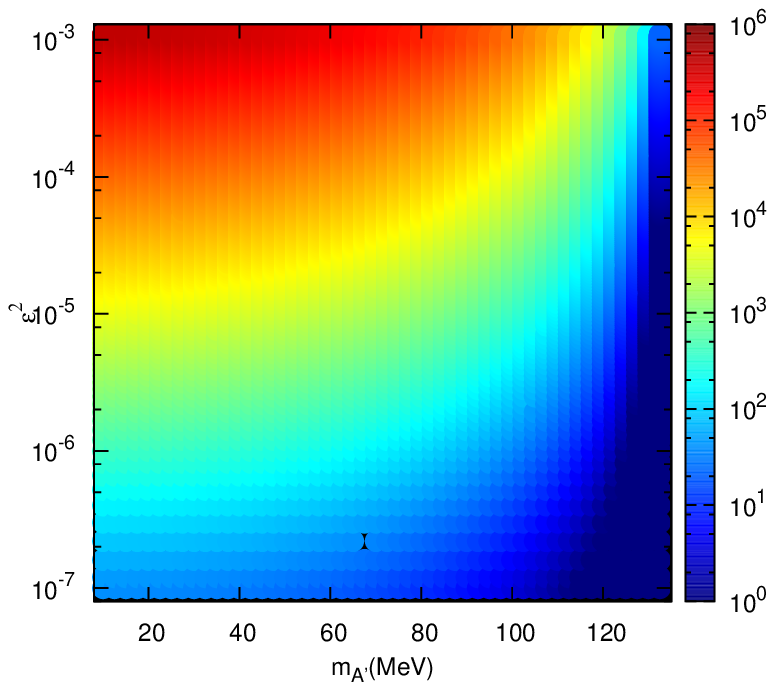,width=5.3cm}} & 
 {\psfig{figure=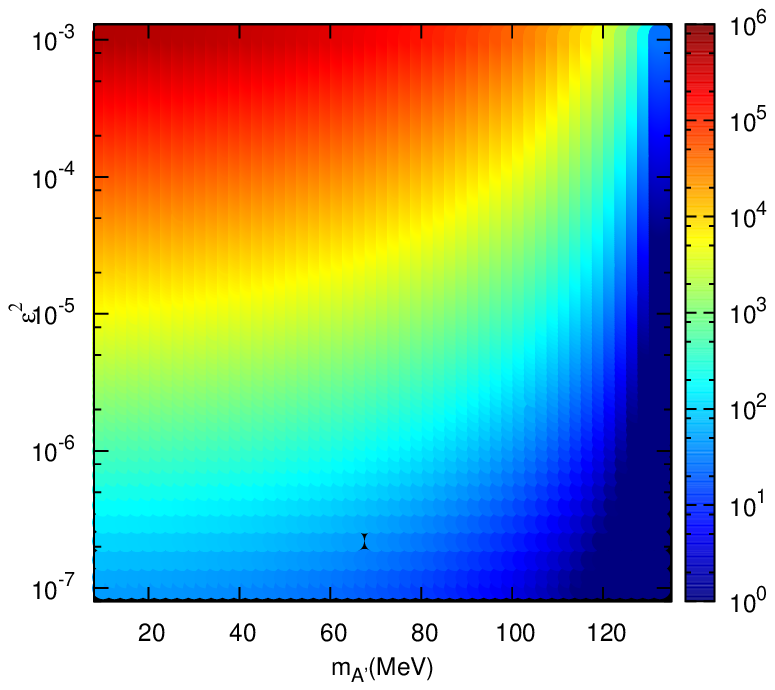,width=5.3cm}} & {\psfig{figure=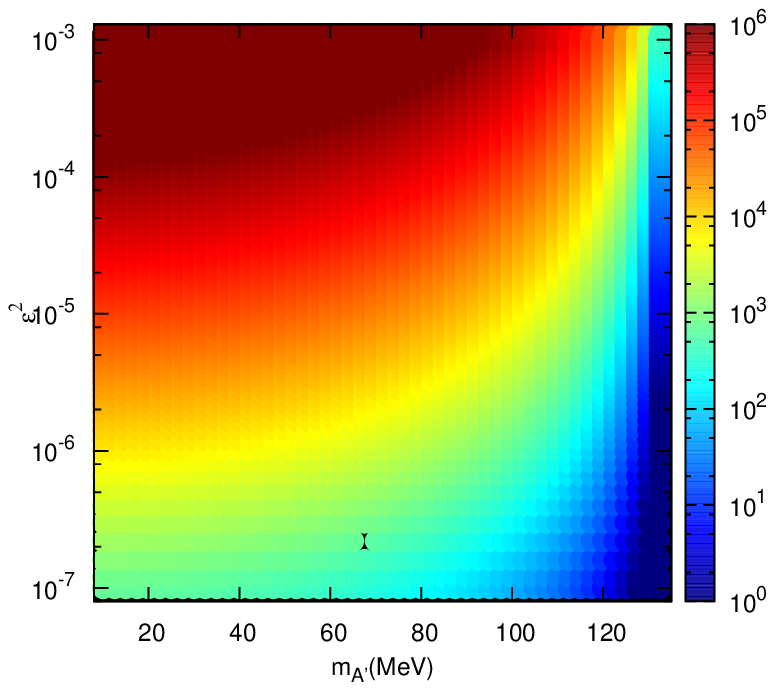,width=5.3cm}} \\
\hspace{-2.2cm} {\psfig{figure=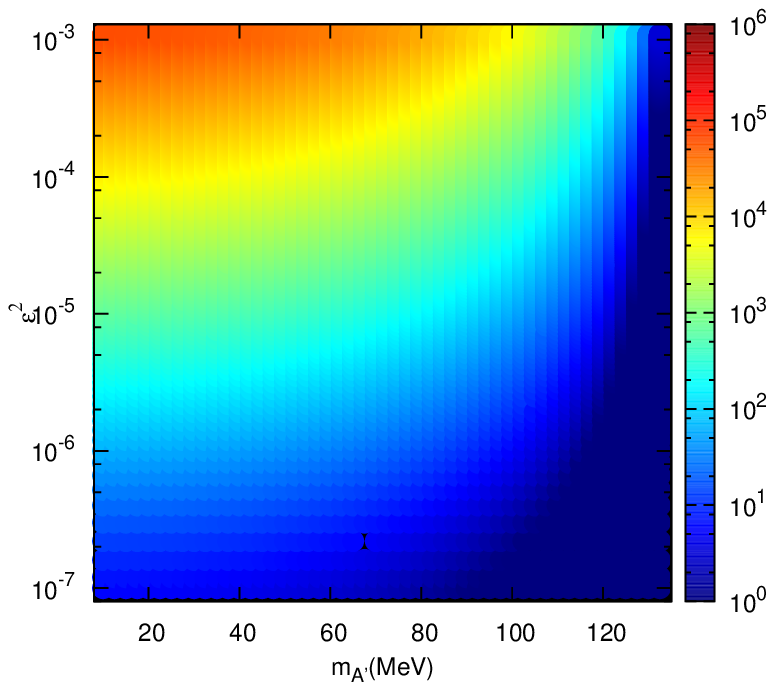,width=5.3cm}} &  {\psfig{figure=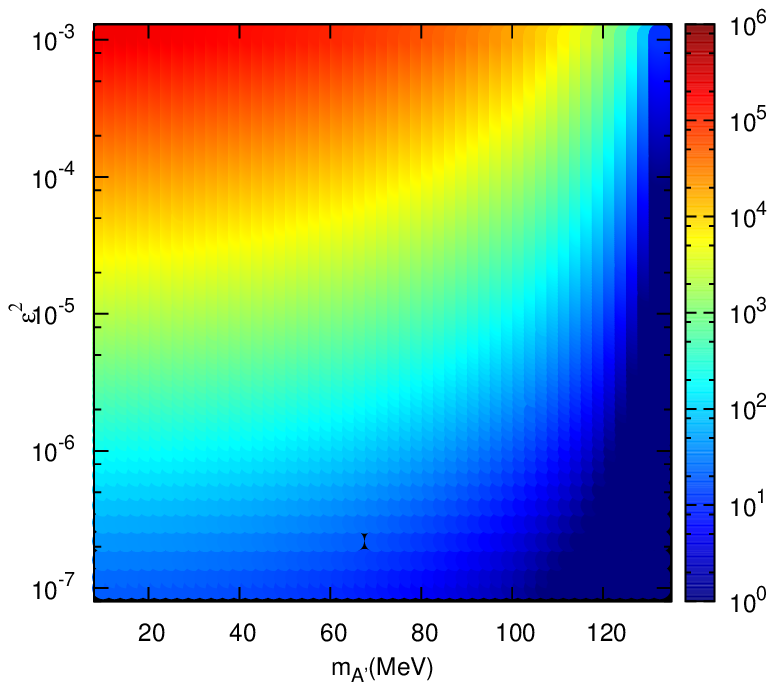,width=5.3cm}} & 
 {\psfig{figure=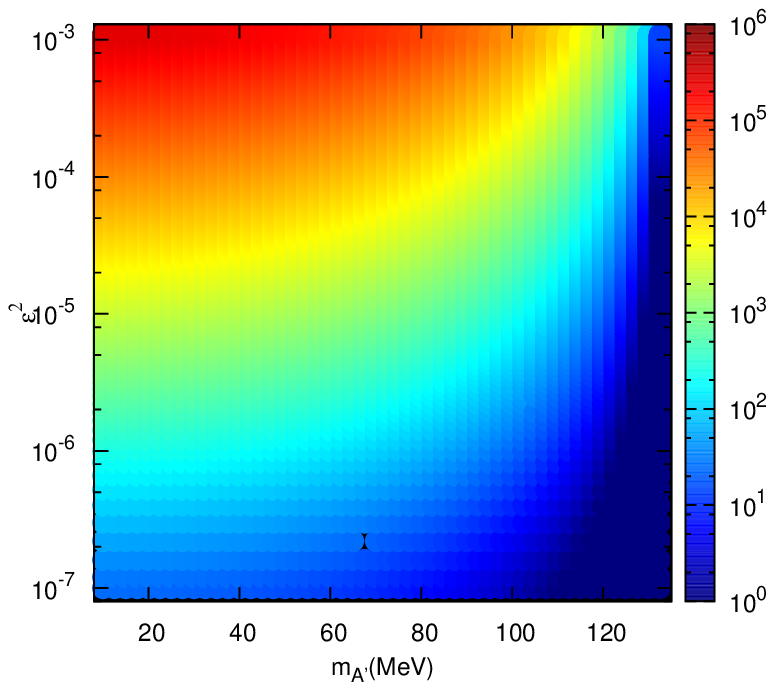,width=5.3cm}} & {\psfig{figure=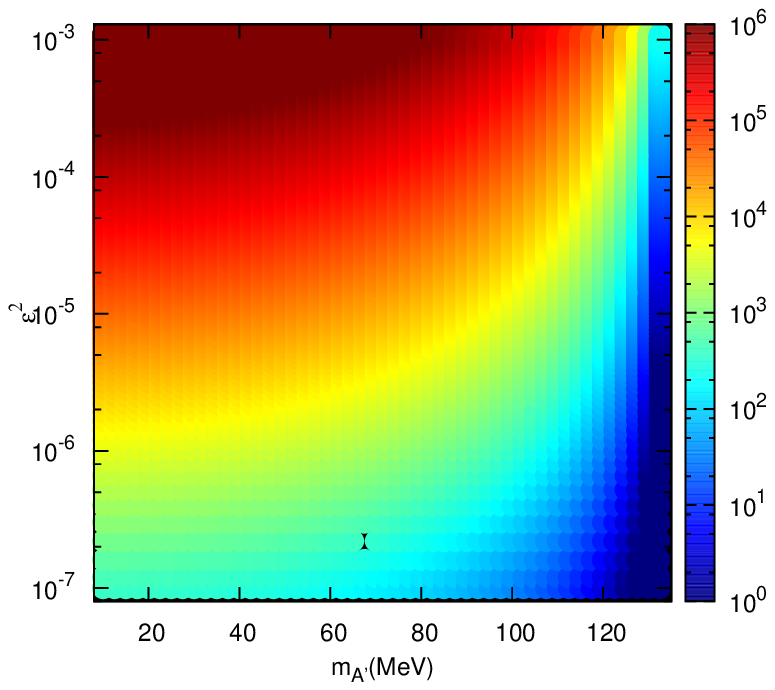,width=5.3cm}} 
\end{tabular}                                                                                                                       
\caption{Two -- dimensional ($m_{A^{\prime}},\epsilon^2$) distribution for the number of events per year in $PbPb$ collisions at the LHC, HL -- LHC, HE -- LHC and FCC   considering the central (upper panels) and forward (lower panels) rapidity ranges.}
\label{fig:events}
\end{figure}

In what follows we  present our predictions for the number of events expected in ultraperipheral $PbPb$ collisions for the planned center -- of -- mass energies of next run of the LHC, as well for the future High -- Luminosity LHC (HL -- LHC) \cite{hl_lhc},  High -- Energy LHC  (HL -- LHC) \cite{he_lhc} and Future Circular Collider (FCC) \cite{fcc}, which are $\sqrt{s} = 5.5, \, 5.5, \, 10$ and 39 TeV, respectively. Moreover, following Refs. \cite{hl_lhc,he_lhc,fcc}, we will assume 
that the integrated luminosities are ${\cal{L}} = 3.0 /\,10 /\,10 /\,110$ $nb^{-1}$ for LHC / HL -- LHC /  HE -- LHC / FCC. The associated predictions are presented in Fig. \ref{fig:events}, where we present the two -- dimensional ($m_{A^{\prime}},\epsilon^2$) distribution for the number of events per year in $PbPb$ collisions    considering the central (upper panels) and forward (lower panels) rapidity ranges.
For the next run of the LHC, 
 we predict that the number of events will be larger than  $10^3$ for  $\epsilon^2 \ge 4 \times 10^{-6}$ and $m_{A^{\prime}} \le 60 $ MeV. On the other hand, for the FCC, one has that event rates per year will be larger than $10^4$ for  $m_{A^{\prime}} \le 100 $ MeV. Such large values indicate that the search of dark photons from pions produced in UPHIC  will be, in principle, feasible.

Finally, let us summarize our main conclusions. In this exploratory study we have investigated the production of a dark photon from pions produced  in ultraperipheral $PbPb$ collisions at different center -- of -- mass energies. We have used the equivalent photon approximation and considered a realistic model for the nuclear photon flux and for the treatment of the absorptive corrections. We have estimated, for the first time, the 
the associated cross sections and event rates  for the center -- of -- mass and integrated luminosities expected for the LHC, HL -- LHC,  HE -- LHC and FCC. Our main motivation is the possibility of use this process to search by a dark photon. 
 We predict large values for the cross sections and event rates, which indicate that a future experimental analysis of the dark photons from pions in UPHIC is, in principle, feasible. The results presented in this letter strongly motivate  a more detailed analysis, including the cuts usually considered by the experimental collaborations. Such study is currently being  performed.

\begin{acknowledgments}
VPG acknowledge very useful discussions about dark photon search at LHCb with M. S. Rangel.
This work was  partially financed by the Brazilian funding
agencies CNPq, CAPES,  FAPERGS and INCT-FNA (process number 
464898/2014-5).
\end{acknowledgments}

\hspace{1.0cm}

\end{document}